\documentclass[prl,floatfix,aps,twocolumn,nofootinbib,superscriptaddress]{revtex4}
\usepackage{graphicx}
\usepackage{epsfig}

\usepackage{amsmath}
\usepackage{amsfonts}
\usepackage{amssymb}
\usepackage{bm}
\usepackage{mathtools}
\usepackage{xcolor}
\usepackage{subcaption}

\usepackage{hyperref}
\usepackage{dsfont}
\usepackage{lipsum}

\usepackage{tikz}
\usepackage{circuitikz}
\usetikzlibrary{arrows, shapes}
\usepackage{pgfplots}
\pgfplotsset{compat=1.14}

\usepackage[final]{pdfpages}

\newcommand{\change}[1]{\textcolor{black}{#1}}
\newcommand{\changet}[1]{\textcolor{black}{#1}}

\begin{document}
\author{Karel Proesmans}
\affiliation{Niels Bohr International Academy, Niels Bohr Institute, University of Copenhagen, Blegdamsvej 17, 2100 Copenhagen, Denmark}
\affiliation{Hasselt University, B-3590 Diepenbeek, Belgium.}

\title{Precision-dissipation trade-off \changet{and optimal protocols} for driven stochastic systems}

\date{\today}

\begin{abstract}
In this paper, I derive a closed expression for how precisely a small-scaled system can follow a pre-defined trajectory, while keeping its dissipation below a fixed limit. The total amount of dissipation is approximately inversely proportional to the expected deviation from the pre-defined trajectory. The \changet{optimal driving protocol is derived and it is shown that associated time-dependent probability distribution conserves its shape throughout the protocol.} \change{Potential applications are discussed in the context of bit erasure and electronic circuits.}

\end{abstract}

\maketitle

\section{Introduction}
The dynamics of mesoscopic systems are heavily influenced by thermal fluctuations. Controlling those systems generally incurs a thermodynamic cost. Over the last decade, several general bounds on this cost have been derived within the framework of stochastic thermodynamics \cite{seifert2012stochastic,peliti2021stochastic}. For example, the thermodynamic uncertainty relation states that the signal-to-noise ratio of any thermodynamic flux is bounded by the dissipation rate (i.e., entropy production rate) of the system \cite{barato2015thermodynamic,gingrich2016dissipation,proesmans2017discrete,hasegawa2019fluctuation,timpanaro2019thermodynamic,koyuk2019operationally,proesmans2019hysteretic,harunari2020exact,pal2020experimental}; the thermodynamic speed limit states that the speed at which a system can be transferred from an initial to a specific final finite state is bounded by the dissipation rate \cite{aurell2011optimal,aurell2012refined,Sivak2012TDMetrics,shiraishi2018speed,proesmans2020finite,proesmans2020optimal,ito2020stochastic,zhen2021universal,van2022finite,dechant2022minimum}; and the dissipation-time uncertainty relation bounds the total amount of dissipation for any first-passage time of far-from-equilibrium systems \cite{falasco2020dissipation,kuznets2021dissipation,yan2022experimental}. \change{These bounds have lead to several applications, such as assessments of the efficiency of cellular processes and methods to infer the dissipation rate from measurements of thermodynamic fluxes \cite{pietzonka2016universal,gingrich2017inferring,pietzonka2018universal,li2019quantifying,manikandan2021quantitative}.}

The central question of this paper is: 'How precisely can a system follow a pre-defined trajectory while keeping the total \change{expected} amount of dissipation associated with the process below a fixed limit?'. This paper answers this question for general continuous-state Markov systems, by deriving an expression for the minimal expected deviation between the desired trajectory and the actual trajectory, given a fixed amount of entropy production and provided that one has full control over the system. The collection of optimal solutions for different values of total \change{expected} dissipation, also known as a Pareto front, generally has a rather complicated form (cf. Eqs.~\eqref{eq:epsy}-\eqref{eq:yres0} below), but simplifies in the small-deviation limit. \change{In this limit,} the minimal expected deviation from the desired trajectory is inversely proportional to the amount of dissipation in the process. This precision-dissipation trade-off relation and the associated optimal protocol \changet{open several new research directions. For example, in contrast to the existing bounds mentioned above, this relation gives a lower bound for the entropy production in terms of directly experimentally accessible quantities. In this way, one can extend existing applications, such as inference of dissipation rate to arbitrary time-dependent systems.}

The next section of this paper introduces the basic notation and reviews some results of stochastic thermodynamics that will then be used to derive the precision-dissipation trade-off relation and the associated driving protocols. \change{Subsequently, I will show general applications of the framework in information processing and in electric circuits}. The paper ends with a discussion on  potential applications and future research directions.

\section{Stochastic thermodynamics\label{sec:st}}
Throughout this paper, I will focus on \change{$n$-dimensional continuous Markov} systems whose state can be described by a variable $\mathbf{x}=(x_1,x_2,..,x_n)$. The probability, $p(\mathbf{x},t)$, for the system to be in state $\mathbf{x}$ at time $t$ satisfies an overdamped Fokker-Planck equation:
\begin{equation}
    \frac{\partial}{\partial t}p(\mathbf{x},t)=-\nabla\cdot\left(\mathbf{v}(\mathbf{x},t)p(\mathbf{x},t)\right),\label{eq:fp}
\end{equation}
where $\mathbf{v}(\mathbf{x},t)$ is the probability flux, given by,
\begin{equation}
   \change{ \mathbf{v}(\mathbf{x},t)=\frac{D}{k_BT}\left(\mathbf{F}(\mathbf{x},t)-k_BT\nabla\ln\left(p(\mathbf{x},t)\right)\right).}\label{eq:pf}
\end{equation}
Here \change{$\mathbf{F}(\mathbf{x},t)$ is the force field that the system experiences}, $D$ is the diffusion coefficient, $k_B$ is the Boltzmann constant, and $T$ is the temperature of the environment. Throughout this paper, $T$ and $D$ are assumed to be constant.
\change{The Fokker-Planck} equation\change{, Eq.~\eqref{eq:pf},} can describe a broad class of systems including the position of a colloidal particle in a potential energy landscape \cite{blickle2006thermodynamics}, \change{the distribution of electrical charges across the conductors} of a linear electrical circuit \cite{freitas2020stochastic}, the state of a spin system \cite{garanin1997fokker}, or the concentration\change{s} of molecular species in a chemical reaction network \cite{gillespie2000chemical}. \change{Throughout this paper, I will assume that one has full control over the force field at all times, unless specified otherwise. This means that one can construct any time-evolution for the probability distribution $p(\mathbf{x},t)$ (cf. SI, section 'Pareto front').}

The goal of this paper is to \change{determine the optimal time-dependent force field}, such that the state of the system, $\mathbf{x}$, follows a given \change{pre-defined} trajectory $\mathbf{X}(t)$ as closely as possible between an initial and a final time, $t=0$ and $t=t_f$. The expected deviation from $\mathbf{X}(t)$ can be quantified by a function $\epsilon$, defined as
\begin{equation}
    \epsilon=\int^{t_f}_0dt\,\int^{\infty}_{-\infty}d\mathbf{x}\,p(\mathbf{x},t)\left(\mathbf{x}-\mathbf{X}(t)\right)^2\label{eq:eps},
\end{equation}
\change{i.e., the expected squared distance from the pre-defined trajectory integrated over the duration of the protocol.}
\change{Meanwhile, one also wants to minimize the amount of dissipation. Stochastic thermodynamics dictates that the average amount of entropy dissipated throughout the process is given by \cite{seifert2012stochastic}}
\begin{equation}
    \Delta_iS=\int^{t_f}_0dt\,\int^{\infty}_{-\infty}d\mathbf{x}\,\frac{k_B}{D}\mathbf{v}(\mathbf{x},t)^2p(\mathbf{x},t).
    \label{eq:sig}
\end{equation}
\change{Note that this expected amount of dissipation is always positive, in accordance with the second law of thermodynamics.}

\section{Precision-dissipation trade-off\label{sec:mr}}
The central goal of this paper will be to minimize the expected deviation, $\epsilon$, while also minimizing the \change{expected} amount of dissipation, $\Delta_i S$, by choosing an optimal protocol for the \change{force field, $\mathbf{F}(\mathbf{x},t)$}.
$\epsilon$ is generally minimized by immediately forcing the probability distribution to be peaked around $X(t)$, which leads to a diverging amount of dissipation. Meanwhile, the total dissipation can be set to zero, by setting \change{$\mathbf{F}(\mathbf{x},t)=k_BT\nabla\ln(p(\mathbf{x},0))$} at all times. \change{One can use Eqs.~\eqref{eq:fp} and \eqref{eq:sig} to show that this force fields leads to a stationary state with zero dissipation, but in this case $\epsilon$ will generally be large.}
\change{In other words, the minimization of $\epsilon$ and $\Delta_i S$, are mutually incompatible and there is no unique optimal trajectory.  Therefore, the focus of this paper will be on minimizing the expected deviation $\epsilon$ for a fixed amount of dissipation $\Delta_iS$.} Such a minimum is known as a Pareto-optimal solution \cite{ishizaka2013multi,Solon2018}. The collection of all Pareto-optimal solutions is known as the Pareto front. For any point on the Pareto front, one can only lower the expected deviation, $\epsilon$, by increasing the amount of dissipation and vice versa.
\change{In the SI (section 'Pareto front') I show that the Pareto front can generally be written as a parametric set of equations:}
\change{\begin{eqnarray}
\epsilon&=&\left\langle\int^{t_f}_0dt\,\left(a_\lambda(t)\mathbf{x}+\mathbf{b}_\lambda(t)-\mathbf{X}(t)\right)^2\right\rangle_0,\label{eq:epsy}\\
\Delta_i S&=&\frac{k_B}{D}\left\langle\int^{t_f}_0dt\,\left(\frac{\partial}{\partial t}a_\lambda(t)\mathbf{x}+\frac{\partial}{\partial t}\mathbf{b}_\lambda(t)\right)^2\right\rangle_0,\label{eq:sigy}\nonumber\\
\end{eqnarray}}

\change{with}
\change{\begin{eqnarray}
a_\lambda(t)&=&\frac{\cosh\left(\lambda(t_f-t)\right)}{\cosh(\lambda t_f)},\nonumber\\
\mathbf{b}_\lambda(t)&=&\frac{\lambda\sinh(\lambda t)}{\cosh(\lambda t_f)}\int^{t_f}_0d\tau\,\cosh(\lambda\tau)\mathbf{X}(t_f-\tau)\nonumber\\&&-\lambda\int^{t}_0d\tau\,\sinh(\lambda\tau)\mathbf{X}(t-\tau)\label{eq:yres0},
\end{eqnarray}}
\change{and $\langle.\rangle_{0}$ stands for the average taken over the initial probability distribution $p(\mathbf{x},0)$, $\left\langle g(\mathbf{x})\right\rangle=\int^{\infty}_{-\infty}d\mathbf{x}p(\mathbf{x},0)g(\mathbf{x})$, for any test function $g(\mathbf{x})$. The full Pareto front can be found by varying $\lambda$ between zero ($\Delta_iS=0$) and infinity ($\epsilon=0$).}

\change{Remarkably, the Pareto front only depends on the initial state of the system through its first two moments, $\langle \mathbf{x}^2\rangle_0=\int^{\infty}_{-\infty}d\mathbf{x}\,p(\mathbf{x},0)\mathbf{x}^2$ and $\langle \mathbf{x}\rangle_0=\int^{\infty}_{-\infty}d\mathbf{x}\,p(\mathbf{x},0)\mathbf{x}$. This shows that the Pareto front for very distinctive problems can have exactly the same shape. In particular, one can map any Pareto front on that of a Gaussian system with the same initial average and variance.}

\change{The main strength of the closed set of equations, Eqs.~\eqref{eq:epsy}-\eqref{eq:yres0}, is its broad applicability, but its complicated shape makes it hard to get an immediate physical intuition. In most relevant applications, however, one is primarily interested in reaching a very high level of precision, i.e., very small $\epsilon$. In this limit, the expression for the Pareto front simplifies to (c.f., SI section 'High-Precision limit')}
\change{\begin{eqnarray}
\Delta_i S\sim\frac{k_B\left(\left\langle(x-X(0))^2\right\rangle_0+\sum_{i\in\{\textrm{jumps}\}}\left(\Delta \mathbf{X}_i\right)^2\right)^2}{4D\epsilon},\label{eq:diseps}
\end{eqnarray}}
\change{where $\{\textrm{jumps}\}$ stands for the collection of discontinuities in the protocol, $\lim_{t\rightarrow t_i^+}\mathbf{X}(t)-\lim_{t\rightarrow t_i^-}\mathbf{X}(t)\equiv\Delta\mathbf{X}_i\neq 0$}. Therefore, one can conclude that in the high-precision limit, the minimal \change{expected} amount of dissipation is inversely proportional to the \change{expected} deviation from the desired trajectory. 

So far, \change{I have focused on the expression of the Pareto front, but it is also possible to obtain the associated optimal protocols for the force field and the probability distribution. Firstly, the optimal time-dependent force field is given by (c.f., SI, section 'Optimal Protocols')}
\change{\begin{eqnarray}
 \mathbf{F}(\mathbf{x},t)&=&-\nabla U(\mathbf{x},t),\nonumber\\
 U(\mathbf{x},t)&=&\frac{k_BT}{D}\left(a_{\lambda}(t)\frac{\partial}{\partial t}\left(\frac{\left(\mathbf{x}-\mathbf{b}_\lambda(t)\right)^2}{2a_{\lambda}(t)}\right)\right.\nonumber\\&&\qquad\quad\qquad\left.-D\ln p\left(\frac{\mathbf{x}-\mathbf{b}_{\lambda}(t)}{a_{\lambda}(t)},0\right)\right).\label{eq:fopt}
\end{eqnarray}
This force field is of a gradient form, i.e., the optimal protocol only involves a conservative energy landscape and non-conservative forces will generally not improve the precision of the driving without inducing extra dissipation. This is in stark contrast to discrete-state systems, where non-conservative forces are generally necessary to minimize dissipation \cite{remlein2021optimality}. Eq.~\eqref{eq:fopt} also reveals the level of control needed to reach the Pareto front: the optimal energy landscape is the sum of a time-dependent harmonic oscillator and an energy landscape that has the same shape as the equilibrium energy landscape at $t=0$, $-k_BT\ln p(\mathbf{x},0)$.} \change{This means that if the system is initially in a Gaussian state, the optimal energy landscape, Eq.~\eqref{eq:fopt}, corresponds to an harmonic oscillator at all times.} \change{This expression also gives a clear interpretation to the functions $a_{\lambda}(t)$ and $\mathbf{b}_{\lambda}(t)$: $a_{\lambda}(t)$ is a decreasing function, independent of the target trajectory $\mathbf{X}(t)$, which leads to a tightening of the energy landscape throughout the protocol, while $\mathbf{b}_\lambda(t)$ determines the positional shift of the energy landscape.}

\change{It is also possible to calculate the probability distribution associated with the state of the system at all times (c.f., SI, section 'Optimal Protocols'):}
\begin{equation}
    p(\mathbf{x},t)=\frac{p\left(\frac{\mathbf{x}-\mathbf{b}_{\lambda}(t)}{a_{\lambda}(t)},0\right)}{a_{\lambda}(t)^n}.\label{eq:ptfin}
\end{equation}
In other words, the protocol that minimizes $\epsilon$ for a given value of $\Delta_iS$ conserves the shape of the probability distribution associated with the state of the system at all times. \change{This result is in agreement with the aforementioned interpretation that $a_\lambda(t)$ is responsible for the narrowing of the distribution while $\mathbf{b}_\lambda(t)$  leads to a positional shift.}

\section{Applications} \label{sec:exa}
\begin{figure}
    \begin{subfigure}{.5\textwidth}
  \centering
\includegraphics[scale=0.5]{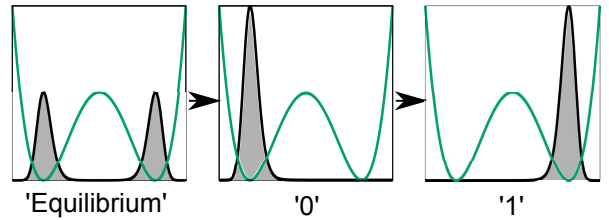}
  \caption{\change{Erasure plus bit-flip of a bit that is initially in equilibrium. The probability distribution is shown in black, while the potential energy landscape at time $t=0$, $U_0(x)$, is shown in green.}}
  \label{fig:sfig1}
\end{subfigure}%
\\
\begin{subfigure}{.5\textwidth}
  \centering
    \includegraphics{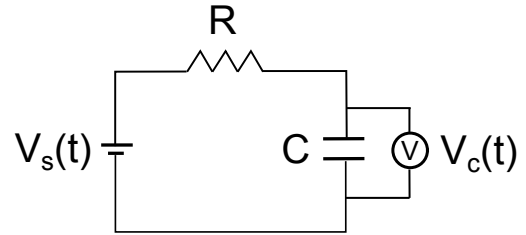}
    \caption{Electrical circuit with time-dependent voltage source $V_s(t)$, resistor $R$ and capacitor $C$.}
    \label{fig:circuit}
    \end{subfigure}
    \caption{\change{Schematic drawings of the applications discussed in the main text.}}
    \label{fig:rep}
\end{figure}
\change{The general bound and optimal protocols derived above can be applied to a broad class of systems. To illustrate this, I will look at two examples (cf.~Fig.~\ref{fig:rep}): information processing, where this framework allows to optimize arbitrary complicated bit operations, and electronic circuits, which serve as an ideal setting to check how well systems under limited control can approach the Pareto front.}

\begin{figure}
    \begin{subfigure}{.5\textwidth}
  \centering
\includegraphics[scale=0.5]{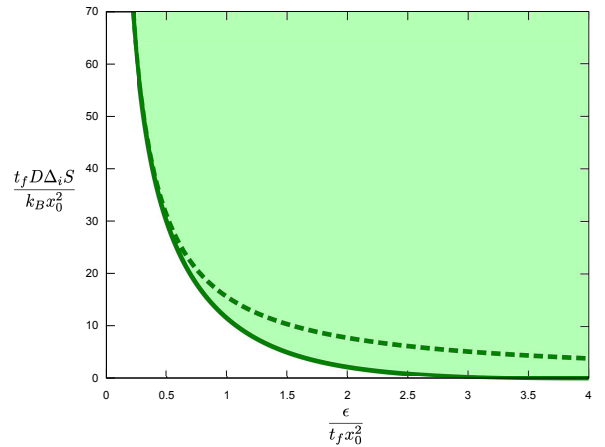}
  \caption{\change{Pareto front for erasure plus bit-flip, with $c=\sqrt{2}$, $E_0=2k_BT$ and $c=\sqrt{2}$.} \change{The dashed red line corresponds to Eq.~\eqref{eq:diseps}}.}
  \label{fig:sfig3}
\end{subfigure}%
\\
\begin{subfigure}{.5\textwidth}
  \centering
    \includegraphics[scale=0.5]{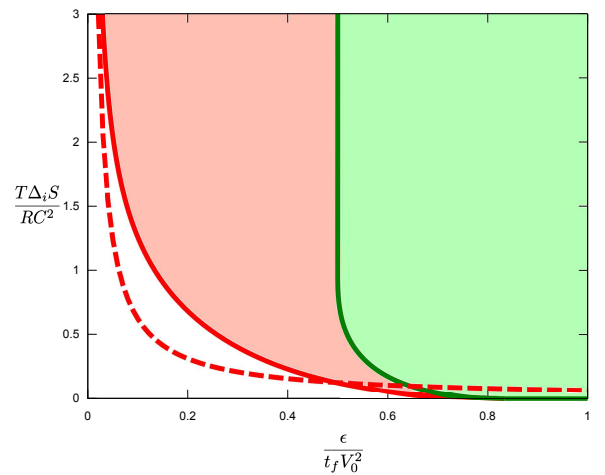}
    \caption{Pareto front for an electrical circuit with time-dependent voltage source $V_s(t)$, resistor $R$ and capacitor $C$, with $t_f=RC/3$ and $CV_0^2=2k_BT$.  The green region corresponds to the values of $\epsilon$ and $\Delta_iS$ that can be reached by only controlling the voltage source, whereas the red region corresponds to the theoretical Pareto front under full control. \change{The dashed red line corresponds to Eq.~\eqref{eq:diseps}}.}
    \label{fig:circuit2}
    \end{subfigure}
     \caption{\change{Pareto fronts for (a) information erasure and (b) the electrical circuit, as depicted in Fig.~\ref{fig:rep}.}}
    \label{fig:circuitfull}
\end{figure}

\change{Computational bits can be modelled using a one-dimensional Fokker-Planck equation with a double-well potential \cite{berut2012experimental,jun2014high},}
\change{\begin{eqnarray}
    \frac{\partial}{\partial t}p(x,t)&=&\frac{D}{k_BT}\frac{\partial}{\partial x}\left(p(x,t)\frac{\partial}{\partial x}U_0(x)+k_BT\frac{\partial}{\partial x}p(x,t)\right),\nonumber\\
    U_0(x)&=&E_0\left(\left(\frac{x}{x_0}\right)^4-2c^2\left(\frac{x}{x_0}\right)^2\right),
\end{eqnarray}}
\change{where $E_0$, $x_0$ and $c_0$ are free parameters.}
\change{The bit is than said to be in state $0$ if $x<0$ and in state $1$ if $x>0$. In equilibrium, the bit is equally likely to be in state $0$ and to be in state $1$, $p(x<0)=p(x>0)=1/2$. If one wants to erase the bit to state $0$, $p(x<0)=1$, one needs to perform an amount of work, $W$, to the system by modulating $U(x,t)$. The expectation value of $W$ is bounded bounded by Landauer's limit, $\left\langle W\right\rangle\geq k_BT\ln2$. Over the last decade, several methods have been derived to extend Landauer's principle to finite-time processes \cite{aurell2011optimal,proesmans2020finite,proesmans2020optimal,proesmans2021erasing,zulkowski2014optimal,zulkowski2015optimal,boyd2022shortcuts}, where one can show that one needs to put in an extra amount of work, corresponding to the dissipation. With the framework derived in this paper it is possible to extend these results to more complicated bit operations. In particular, I will focus on the precision-dissipation trade-off associated with erasing a bit to state $0$ and subsequently flipping the bit to state $1$, c.f.~Fig.~\ref{fig:sfig1}. This corresponds to}
\change{
\begin{equation}
X(t)=\begin{cases}-c_0x_0 & t<\frac{t_f}{2}\\
c_0x_0 & t>\frac{t_f}{2}
\end{cases}\label{eq:Xtinf}
\end{equation}
and $p(x,0)$ corresponds to the equilibrium distribution associated with $U_0(x)$. The resulting Pareto front is shown in Fig.~\ref{fig:sfig3} (c.f., SI section 'Applications' for detailed calculations). The explicit protocols in the low-deviation ($\epsilon=0.5$) and the high-deviation ($\epsilon=2$) limit are shown in the supplementary videos. Furthermore, SI Fig.~\ref{fig:aandb} in the supplemental materials shows $a_{\lambda}(t)$ and $b_{\lambda}(t)$ for different levels of precision. One can verify that the positional shift, $b_{\lambda}(t)$ follows $X(t)$ closer at higher precision and mainly deviates around the jump in $X(t)$. Meanwhile the probability distribution tightens exponentially fast, as illustrated by the decay of $a_\lambda(t)$.
}

\change{There are many systems, where one does not have full control over the driving protocol. This can make it impossible to implement the optimal protocol, Eq.~\eqref{eq:diseps}, and saturate the bound, Eqs.~\eqref{eq:epsy}-\eqref{eq:yres0}. It is not a priori clear how close the Pareto front under limited control is to the one under full control. To test this, I will now turn to }the electronic circuit shown in Fig.~\ref{fig:circuit}, where an observer controls a time-dependent voltage source $V_s(t)$ connected to a resistor, with resistance $R$, and a capacitor, with capacitance $C$. One can then use the time-dependent voltage source to make sure that voltage over the capacitor, $v_C$, follows a pre-defined trajectory. For small-scaled systems, this voltage will generally fluctuate due to thermal noise. Control over $V_s(t)$ does not allow for any arbitrary force field $F(v_c,t)$, as will be shown below. Therefore, the minimal deviation for a given \change{amount of} dissipation will not saturate the Pareto front.

The voltage fluctuations associated with thermal noise are given by the Johnson-Nyquist formula \cite{freitas2020stochastic}. In this case, one can show that the voltage over the capacitor satisfies:
\begin{multline}
    \frac{\partial}{\partial t}p(v_c,t)=-\frac{\partial}{\partial v_c}\left(\frac{\left(V_s(t)-v_c\right)}{RC}p(v_c,t)\right)\\+\frac{k_BT}{RC^2}\frac{\partial^2}{\partial v_c^2}p(v_c,t).\label{eq:fpvc}
\end{multline}
This corresponds to a Fokker-Planck equation, similar to Eq.~\eqref{eq:fp}, with
\change{$F(v_c,t)=C(V_s(t)-v_c$)} and $D=k_BT/(RC^2)$.
The voltage source can now be used to apply a time-dependent voltage over the capacitor. \change{Here, I will focus on a protocol where one tries to charge the capacitor 
\begin{equation}
    V_T(t)=\frac{V_0 t}{t_f}.\label{eq:Xelec}
\end{equation}}
Initially, the capacitor is assumed to be in equilibrium with the voltage source,
\begin{equation}
    p(v_c,0)=\sqrt{\frac{k_BT}{2\pi C}}e^{-\frac{Cv_c^2}{2k_BT}}.
\end{equation}
With this boundary conditions, one can write a general expression for the probability distribution at all times:
\begin{equation}
    p(v_c,t)=\sqrt{\frac{k_BT}{2\pi C}}e^{-\frac{C\left(v_c-V_C(t)\right)^2}{2k_BT}},
\end{equation}
with
\begin{equation}
    V_C(t)=\int_0^td\tau\,\frac{V_s(\tau)e^{\frac{\tau-t}{CR}}}{CR}.\label{eq:vcres}
\end{equation}
This expression can be verified by plugging it in into Eq.~\eqref{eq:fpvc}.
One can use the general framework for thermodynamics of electronic circuits to calculate the total amount of dissipation during the process \cite{freitas2020stochastic}. This gives an expression that corresponds exactly to Eq.~\eqref{eq:sig}. Furthermore, the precision is defined in the same way as in Eq.~\eqref{eq:eps}, with $X(t)=V_T(t)$.

Both the theoretical precision-dissipation Pareto front under full control, Eq.~\eqref{eq:epsy}-\eqref{eq:yres0} and the Pareto front when one only has control over the voltage-source are calculated explicitly in the SI, \change{(sections 'Applications' and 'Limited control' respectively).} 
The Pareto fronts are shown in Fig.~\ref{fig:circuit2}. The green region corresponds to the values of $\epsilon$ and $\Delta_iS$ that can be reached by controlling the voltage source, whereas the red region corresponds to values of $\epsilon$ and $\Delta_iS$ that can only be reached under full control. One can verify from Fig.~\ref{fig:circuit2} that the high-precision region cannot be reached by only controlling the voltage source.

\change{The supplemental videos show optimal protocols both under full control (for $\epsilon=0.05$ and $\epsilon=0.5$) and under limited control (for $\epsilon=0.5$ and $\epsilon=0.6$). One can see that the full-control protocol primarily focuses on avoiding big fluctuations while the limited control protocol focuses more on optimizing the average value. Furthermore, one can see that the optimal protocol mainly deviates from $X(t)$ at the end of the protocol. This can also be seen in SI Fig.~\ref{fig:aandb}.}

\section{Discussion\label{sec:dis}}
In conclusion, this paper derives a general expression for the minimal thermodynamic cost associated with following a pre-defined trajectory at a given precision. \change{The results show that the expected deviation from the pre-defined trajectory is approximately inversely proportional to the amount of dissipation. The general bound holds for all systems that can be described by an overdamped Langevin system. This means that the results can be applied to a broad class of biological and chemical systems, alongside the examples in information processing and electronic systems discussed above.}

This work opens up several potential directions for future research. \change{One particularly interesting application would be to use the Pareto front to infer a lower bound on the amount of entropy production from experimental measurements of the precision. Indeed, by calculating $\epsilon$ with respect to any choice of $X(t)$, one can use the Pareto front derived in this paper to infer a lower bound on the dissipation rate of the experimental system. In contrast to other existing methods used to infer the dissipation rate of experimental systems \cite{roldan2010estimating,martinez2019inferring,li2019quantifying,ehrich2021tightest}, the bound derived in this paper applies to systems with arbitrary time-dependent driving. The accuracy of this method could be tested on a broad range of experimental systems, including micro-electronic systems similar to the example discussed in this paper \cite{garnier2005nonequilibrium}, or colloidal particles trapped by optical tweezers \cite{ciliberto2017experiments,kumar2022anomalous}.} \change{Another interesting application would be to use the the bound can as a quantitative test in how efficient a choice of control parameters is.}

\change{There are also several ways in which the results from this paper can be extended. For example, one can use the same methodology to derive lower bounds for the entropy production associated with minimizing other observables.} It might also be possible to extend the results of this paper to discrete state systems and systems with strong quantum effects, using similar ideas, as in the known extensions of the thermodynamic speed limit \cite{shiraishi2018speed}. It would also be interesting to compare the results of this paper with known Pareto fronts under limited control, which are known to exhibit phase-transitions \cite{Solon2018}. 

\section*{Acknowledgment}
I thank John Bechhoefer for interesting discussions on the manuscript. This project has received funding from the European Union’s Horizon 2020 research and innovation program under the Marie Sklodowska-Curie grant agreement No. 847523 ‘INTERACTIONS’ and from the Novo Nordisk Foundation (grant No. NNF18SA0035142 and NNF21OC0071284).

\newpage
a
\appendix
\newpage
\widetext
\section{\large{Supplemental Information}}

\section{Pareto front}
\change{The Pareto front can be found using Lagrangian techniques. To obtain the minimal value of $\epsilon$ for fixed $\Delta_i S$, one needs to minimize the Lagrangian}
\begin{eqnarray}
    \mathcal{L}= \epsilon+\lambda_0\Delta_i S,\label{eq:lan1}
\end{eqnarray}
with respect to $\mathbf{F}(\mathbf{x},t)$. Here $\lambda_0$ is a Lagrange multiplier, which can be used to fix $\Delta_iS$. \change{This minimisation seems highly non-trivial as $p(\mathbf{x},t)$ depends on $\mathbf{F}(\mathbf{x},t)$ in a non-trivial way, c.f., Eqs.~\eqref{eq:fp}-\eqref{eq:pf}. One can do this minimization by introducing a transport map, $\mathbf{y}(\mathbf{x},t)$, defined through
\begin{equation}
    \frac{\partial}{\partial t}\mathbf{y}(\mathbf{x},t)=\mathbf{v}(\mathbf{y}(\mathbf{x},t),t),\qquad \mathbf{y}(\mathbf{x},0)=\mathbf{x}.\label{eq:ydef}
\end{equation}
This change of variables is mathematically equivalent to changing from an Eulerian to a Lagrangian picture in fluid mechanics. One can use this equivalence to show that \cite{benamou2000computational,batchelor2000introduction}
\begin{eqnarray}
\int_{-\infty}^{\infty}d\mathbf{x}\, g(\mathbf{x},t)p(\mathbf{x},t)&=&\int_{-\infty}^{\infty}d\mathbf{x}\, g(\mathbf{y}(\mathbf{x},t),t)p(\mathbf{x},0),\label{eq:hyrel1}\\
\int_{-\infty}^{\infty}d\mathbf{x}\,\mathbf{v}(\mathbf{x},t) g(\mathbf{x},t)p(\mathbf{x},t)&=&\int_{-\infty}^{\infty}d\mathbf{x}\,\frac{\partial}{\partial t}\mathbf{y}(\mathbf{x},t) g(\mathbf{y}(\mathbf{x},t),t)p(\mathbf{x},0),\label{eq:hyrel2}
\end{eqnarray}
for any test function $g(\mathbf{x},t)$. This equivalence to fluid dynamics also implies that $\mathbf{y}(\mathbf{x},t)$ is generally invertable with respect to $\mathbf{x}$ \cite{benamou2000computational,batchelor2000introduction}.
Eq.~\eqref{eq:hyrel1} can be used to obtain $p(\mathbf{x},t)$, by setting $g(\mathbf{x},t)=\delta(\mathbf{x}-\mathbf{y}(\mathbf{x}_0,t))$ with $\mathbf{x}_0$ an arbitrary constant:
\begin{equation}
    p(\mathbf{y}(\mathbf{x}_0,t),t)=\frac{p\left(\mathbf{x}_0,0\right)}{\det\left(\nabla\mathbf{y}(\mathbf{x}_0,t)\right)}.\label{eq:ptres}
\end{equation}
Furthermore, one can determine the force field associated with this time-evolution. Firstly, using Eq.~\eqref{eq:ydef} for $\mathbf{x}=\mathbf{y}^{-1}(\mathbf{x}_0,t)$, one has
\begin{equation}
    \frac{\partial}{\partial t}\mathbf{y}(\mathbf{y}^{-1}(\mathbf{x}_0,t),t)=\mathbf{v}(\mathbf{x}_0,t).
\end{equation}
Plugging this in into Eq.~\eqref{eq:pf} gives
\begin{equation}
    \mathbf{F}(\mathbf{x}_0,t)=\frac{k_BT}{D}\frac{\partial}{\partial t}\mathbf{y}(\mathbf{y}^{-1}(\mathbf{x}_0,t),t)+k_BT{\nabla\ln\left(p(\mathbf{x}_0,t)\right)},\label{eq:fopt2}
\end{equation}
where $\mathbf{y}^{-1}(\mathbf{x_0},t)$ is the inverse of $\mathbf{y}$ with respect to $\mathbf{x}$.
Therefore, fixing $\mathbf{y}(\mathbf{x},t)$ uniquely defines the force field $\mathbf{F}(\mathbf{x},t)$ and the associated probability distribution $p(\mathbf{x},t)$. Meanwhile, one can use Eqs.~\eqref{eq:pf} and \eqref{eq:ydef} to show that $\mathbf{y}(\mathbf{x},t)$ is determined uniquely for a given force field $\mathbf{F}(\mathbf{x},t)$. Optimizing the Lagrangian, Eq.~\eqref{eq:lan1}, with respect to $\mathbf{F}(\mathbf{x},t)$ is therefore, equivalent to optimizing it with respect to $\mathbf{y}(\mathbf{x},t)$.}

\change{From this analysis, one can also verify that it is possible to create any time-evolution for the probability distribution $p(\mathbf{x},t)$ as stated in the main text. Indeed, for any probability distribution, one can find a $\mathbf{y}(\mathbf{x},t)$ that satisfies Eq.~\eqref{eq:ptres} \cite{benamou2000computational,villani2003topics}. The force field leading to this probability distribution is then given by Eq.~\eqref{eq:fopt2}.}

\change{Using these coordinates, the Lagrangian minimization simplifies to
\begin{equation}
\min_{\mathbf{F}(\mathbf{x},t)}\mathcal{L}=\min_{\mathbf{y}(\mathbf{x},t)}\int^{\infty}_{-\infty}d\mathbf{x}\,\int^{t_f}_0dt\,p(\mathbf{x},0)\left(\left(\mathbf{y}(\mathbf{x},t)-\mathbf{X}(t)\right)^2+\frac{k_B\lambda_0 }{D}\left(\frac{\partial}{\partial t}\mathbf{y}(\mathbf{x},t)\right)^2\right).\label{eq:landef}
\end{equation}
The associated Euler-Lagrange equation is given by
\begin{equation}
    \frac{\partial^2}{\partial t^2}\mathbf{y}(\mathbf{x},t)=\frac{D}{k_B\lambda_0}\left(\mathbf{y}(\mathbf{x},t)-\mathbf{X}(t)\right).\label{eq:yeq}
\end{equation}
This equation can be solved for $\mathbf{y}(x,t)$ for any choice of the boundary condition $\mathbf{y}_f(\mathbf{x})\equiv \mathbf{y}(\mathbf{x},t_f)$, 
\begin{equation}
    \mathbf{y}(\mathbf{x},t)=\left(\mathbf{y}_f(\mathbf{x})+\mathcal{X}_\lambda(t_f)\right)\frac{\sinh\left(\lambda t\right)}{\sinh\left(\lambda t_f\right)}\\+\mathbf{x}\frac{\sinh\left(\lambda(t_f-t)\right)}{\sinh\left(\lambda t_f\right)} -\mathcal{X}_\lambda(t)\label{eq:yres}
\end{equation}
where I introduced $\lambda=\sqrt{D/(k_B\lambda_0)}$, and
\begin{equation}
    \mathcal{X}_\lambda(t)=\lambda\int^t_0d\tau\,\sinh\left(\lambda(t-\tau)\right)\mathbf{X}(\tau),
\end{equation}
for notational simplicity.}

\change{One can fully minimize the Lagrangian, Eq.~\eqref{eq:lan1}, by first calculating its minimum for each value of $\mathbf{y}_f(\mathbf{x})$ and subsequently take the minimum over this collection of all possibilities for $\mathbf{y}_f(\mathbf{x})$. Eq.~\eqref{eq:yres} minimizes the Lagrangian for any fixed choice of $\mathbf{y}_f(\mathbf{x})$. The second minimization can be done by filling in this expression into the Lagrangian and subsequently minimize it with respect to $\mathbf{y}_f(\mathbf{x})$,
\begin{equation}
    \frac{\delta \mathcal{L}}{\delta \mathbf{y}_f(\mathbf{x})}=\mathbf{0}.\label{eq:yfcon}
\end{equation}
One has
\begin{equation}
    \frac{\delta y_i(\mathbf{x},t)}{\delta y_{f,j}(\mathbf{x}_0)}=\delta_{ij}\delta(\mathbf{x}-\mathbf{x}_0)\frac{\sinh\left(\lambda t\right)}{\sinh\left(\lambda t_f\right)},
\end{equation}
and
\begin{equation}
    \frac{\delta \frac{\partial}{\partial t}y_i(\mathbf{x},t)}{\delta y_{f,j}(\mathbf{x}_0)}=\delta_{ij}\delta(\mathbf{x}-\mathbf{x}_0)\lambda\frac{\cosh\left(\lambda t\right)}{\sinh\left(\lambda t_f\right)},
\end{equation}
where $i$ and $j$ are indices of $\mathbf{y}(\mathbf{x},t)$ and $\mathbf{y}_f(\mathbf{x})$ respectively, and where $\delta_{ij}$ and $\delta(x-x_0)$ are a Kronecker delta function and a Dirac delta function respectively. One can use this to write
\begin{equation}
    \delta\mathcal{L}=\int^{\infty}_{-\infty}dx\,\int^{t_f}_0dt\,\frac{2p(x,0)}{\sinh(\lambda t_f)}\left(\sinh\left(\lambda t\right)\left(\mathbf{y}(\mathbf{x},t)-\mathbf{X}(t)\right)\cdot\delta\mathbf{y}_f(\mathbf{x})+\frac{\cosh\left(\lambda t\right)}{\lambda}\left(\frac{\partial}{\partial t}\mathbf{y}(\mathbf{x},t)\right)\cdot\delta\mathbf{y}_f(\mathbf{x})\right),
\end{equation}
and therefore Eq.~\eqref{eq:yfcon} is equivalent to}
\begin{equation}
    \int^{t_f}_0dt\,\left(\lambda\sinh\left(\lambda t\right)\left(\mathbf{y}(\mathbf{x},t)-\mathbf{X}(t)\right)+\cosh\left(\lambda t\right)\left(\frac{\partial}{\partial t}\mathbf{y}(\mathbf{x},t)\right)\right)=0,
\end{equation}
or after applying Eq.~\eqref{eq:yeq} to the first term of this equation and doing a partial integration
\begin{equation}
    \left.\frac{\partial}{\partial t}\mathbf{y}(\mathbf{x},t)\right|_{t=t_f}=0.
\end{equation}
Using Eq.~\eqref{eq:yres}, one can see that this is equivalent to
\begin{equation}
    \mathbf{y}_f(x)=\frac{\mathbf{x}}{\cosh\left(\lambda t_f\right)}+\frac{\tanh\left(\lambda t_f\right)}{\lambda}\frac{\partial \mathcal{X}_\lambda(t_f)}{\partial t}-\mathcal{X}_\lambda(t_f).
\end{equation}
One can simplify this further by noting that
\begin{eqnarray}
\frac{\partial \mathcal{X}_\lambda}{\partial t}(t_f)&=&\left.\lambda\frac{\partial}{\partial t}\left(\int^t_0d\tau\,\sinh\left(\lambda(t-\tau)\right)\mathbf{X}(\tau)\right)\right|_{t=t_f}\nonumber\\
&=&\lambda^2\int^{t_f}_0d\tau\,\cosh\left(\lambda(t_f-\tau)\right)\mathbf{X}(\tau).
\end{eqnarray}
Therefore,
\begin{eqnarray}
    \frac{\tanh\left(\lambda t_f\right)}{\lambda}\frac{\partial \mathcal{X}_\lambda(t_f)}{\partial t}-\mathcal{X}_\lambda(t_f)&=&\lambda\int^{t_f}_0d\tau\,\frac{\left(\cosh\left(\lambda(t_f-\tau)\right)\sinh\left(\lambda t_f\right)-\sinh\left(\lambda(t_f-\tau)\right)\cosh\left(\lambda t_f\right)\right)\mathbf{X}(\tau)}{\cosh\left(\lambda t_f\right)}\nonumber\\
    &=&\frac{\lambda\int^{t_f}_0d\tau\,\sinh\left(\lambda\tau\right)\mathbf{X}(\tau)}{\cosh\left(\lambda t_f\right)},
\end{eqnarray}
and,
\begin{equation}
    \mathbf{y}_f(\mathbf{x})=\frac{\mathbf{x}+\lambda\int^{t_f}_0d\tau\,\sinh\left(\lambda\tau\right)\mathbf{X}(\tau)}{\cosh\left(\lambda t_f\right)}.
\end{equation}
\change{Using Eq.~\eqref{eq:yres},
\begin{eqnarray}
   \mathbf{y}(\mathbf{x},t)&=&\mathbf{x}\left(\frac{\sinh\left(\lambda(t_f-t)\right)\cosh(\lambda t_f)+\sinh(\lambda t)}{\sinh\left(\lambda t_f\right)\cosh(\lambda t_f)}\right)+\frac{\lambda\sinh\left(\lambda t\right)}{\sinh\left(\lambda t_f\right)}\int^{t_f}_0d\tau\,\left(\frac{\sinh(\lambda\tau)}{\cosh(\lambda t_f)}+\sinh(\lambda(t_f-\tau))\right)\mathbf{X}(\tau)-\mathcal{X}_\lambda(t)\nonumber\\
   &=&\frac{\mathbf{x}\cosh\left(\lambda(t_f-t)\right)}{\cosh(\lambda t_f)}+\frac{\lambda\sinh(\lambda t)}{\cosh(\lambda t_f)}\int^{t_f}_0d\tau\,\cosh(\lambda(t_f-\tau))\mathbf{X}(\tau)-\lambda\int^{t}_0d\tau\,\sinh(\lambda(t-\tau))\mathbf{X}(\tau)\nonumber\\
   &=&{a}_\lambda(t)\mathbf{x}+\mathbf{b}_\lambda(t),\label{eq:yresfin}
\end{eqnarray}
where I used
\begin{equation}
    \sinh(\lambda t)=\sinh(\lambda(t_f-(t_f-t)))=\sinh(\lambda t_f)\cosh(\lambda(t_f-t))-\sinh(\lambda(t_f-t))\cosh(\lambda t_f),
\end{equation}
and
\begin{equation}
    \sinh(\lambda\tau)=\sinh(\lambda t_f)\cosh(\lambda(t_f-\tau))-\sinh(\lambda(t_f-\tau))\cosh(\lambda t_f),
\end{equation}
for the first and second term of the second equality respectively, and the definitions of $a_\lambda(t)$ and $\mathbf{b}_{\lambda}(t)$, Eq.~\eqref{eq:yres0}, to get the third equality.
}

\change{This equation, together with Eqs.~\eqref{eq:hyrel1}-\eqref{eq:hyrel2}, and the definitions of $\epsilon$ and $\Delta_iS$, Eqs.~\eqref{eq:eps}-\eqref{eq:sig}, leads to the closed expression for the Pareto front, c.f.~Eqs.~\eqref{eq:epsy}-\eqref{eq:yres0}.}

\section{Optimal protocol}
\change{Eq.~\eqref{eq:yresfin} can also be used to obtain explicit expressions for the time-dependent probability distribution associated with the optimal protocol and the corresponding force field. One can first notice that
\begin{equation}
    \textrm{det}\left(\nabla\mathbf{y}(\mathbf{x}_0,t)\right)=a_{\lambda}(t)^n.
\end{equation}
One can then use Eq.~\eqref{eq:ptres} to show that
\begin{equation}
    p(a_\lambda \mathbf{x}+\mathbf{b}_\lambda,t)=\frac{p(\mathbf{x},0)}{a_{\lambda}(t)^n},
\end{equation}
or
\begin{equation}
    p(\mathbf{x},t)=\frac{p(\frac{\mathbf{x}-\mathbf{b}_{\lambda}(t)}{a_{\lambda}(t)},0)}{a_{\lambda}(t)^n},\label{eq:popt}
\end{equation}
which corresponds to Eq.~\eqref{eq:ptfin} of the main text.}

\change{The optimal force field can be calculated using Eqs.~\eqref{eq:fopt2} and \eqref{eq:popt},
\begin{eqnarray}
    \mathbf{F}(\mathbf{x},t)&=&\frac{k_BT}{D}\frac{\partial}{\partial t}a_{\lambda}(t)\left(\frac{\mathbf{x}-\mathbf{b}_\lambda(t)}{{a}_{\lambda}(t)}\right)+\frac{\partial}{\partial t}\mathbf{b}_\lambda(t)+k_BT\nabla \ln p(\mathbf{x},t)\nonumber\\
    &=&-\frac{k_BT}{D}\nabla\left(a_{\lambda}(t)\frac{\partial}{\partial t}\left(\frac{\mathbf{x}^2-2\mathbf{b}_\lambda(t)\cdot\mathbf{x}}{2a_{\lambda}(t)}\right)-D\ln p\left(\frac{\mathbf{x}-\mathbf{b}_\lambda(t)}{a_\lambda(t)},0\right)\right),
\end{eqnarray}
in agreement with Eq.~\eqref{eq:fopt} of the main text.
}

\section{High-precision limit\label{app:hplimit}}
\begin{figure}
    \centering
    \includegraphics[width=0.8\textwidth]{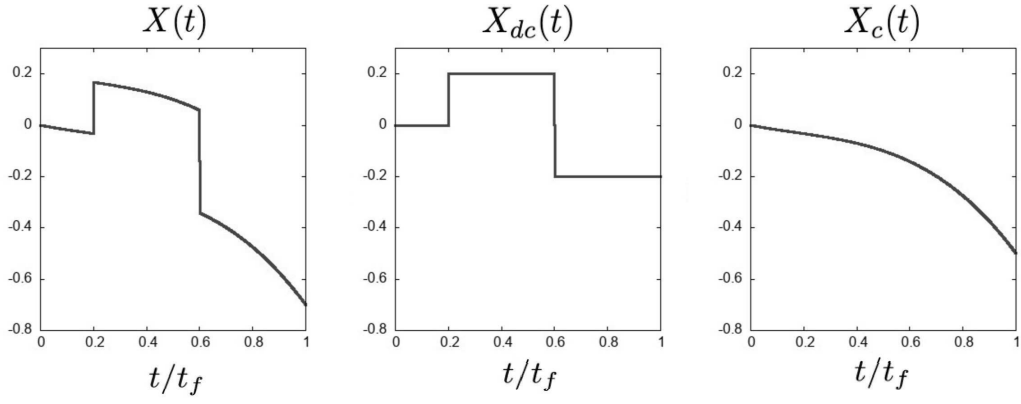}
    \caption{\change{Example of decomposition of $X(t)$ in continuous and discontinuous part ($X_c(t)$ and $X_{dc}(t)$ respectively)}.}
    \label{fig:exXcdc}
\end{figure}
The high-precision limit corresponds to the large-$\lambda$ limit\change{, as can be seen from the Lagrangian Eq.~\eqref{eq:lan1}}.
\change{Throughout this section, I will assume that the target trajectory $\mathbf{X}(t)$ is finite for all values of $t$, but that it might exhibit discontinuities. This means that there exists a unique decomposition}
\change{\begin{equation}
    \mathbf{X}(t)=\mathbf{X}_c(t)+\mathbf{X}_{dc}(t),
\end{equation}}
\change{where $\mathbf{X}_{dc}(t)=\sum_{i}\Delta \mathbf{X}_iH(t-t_i)$, with $H(t)$ the Heaviside function, $t_i$ the time the $i$-th discontinuity in $\mathbf{X}$(t), and $\Delta \mathbf{X}_i=\lim_{t\rightarrow t^{+}_i}\mathbf{X}(t)-\lim_{t\rightarrow t^{-}_i}\mathbf{X}(t)$, and $\mathbf{X}_c(t)$ is a continuous function in $t$. This is illustrated in Fig.~\ref{fig:exXcdc}. One can subsequently define $\mathbf{y}_c(\mathbf{x},t)$ and $\mathbf{y}_{dc}(t)$ as}
\change{\begin{equation}
    \mathbf{y}_{dc}(t)=\frac{\lambda\sinh(\lambda t)}{\cosh(\lambda t_f)}\int^{t_f}_0d\tau\,\cosh(\lambda\tau)\mathbf{X}_{dc}(t_f-\tau)-\lambda\int^{t}_0d\tau\,\sinh(\lambda\tau)\mathbf{X}_{dc}(t-\tau),\qquad \mathbf{y}_c(\mathbf{x},t)=\mathbf{y}(\mathbf{x},t)-\mathbf{y}_{dc}(t).
\end{equation}
}
\change{Being a sum of Heaviside functions, $\mathbf{y}_{dc}(t)$ can be calculated explicitly, using mathematical software:}
\change{
\begin{eqnarray}
    \mathbf{y}_{dc}(t)&=&\mathbf{X}_{dc}(t)+\sum_i\Delta \mathbf{X}_i\left(\frac{\sinh(\lambda t)\sinh(\lambda(t_f-t_i))}{\cosh(\lambda t_f)}-H(t-t_i)\cosh(\lambda(t-t_i))\right).\nonumber\\
\end{eqnarray}
For $t<t_i$ the relevant term in the sum becomes
\begin{eqnarray}
    \frac{\Delta\mathbf{X}_i\sinh(\lambda t)\sinh(\lambda(t_f-t_i))}{\cosh(\lambda t_f)}\approx\Delta\mathbf{X}_i\sinh(\lambda t)e^{-\lambda t_i},
\end{eqnarray}
where I used the fact that in the high-$\lambda$ limit, one has $\sinh(\lambda(t_f-t_i))\approx \exp(\lambda (t_f-t_i))/2$ and $\cosh(\lambda t_f)\approx\exp(\lambda t_f)/2$. Note that the above expression is approximately zero, unless $t$ is very close to $t_i$, in which case $\sinh(\lambda t)\approx\exp(\lambda t)/2$. Therefore, one can conclude that up to a correction that is exponentially small in $\lambda$, one has
\begin{equation}
    \frac{\Delta\mathbf{X}_i\sinh(\lambda t)\sinh(\lambda(t_f-t_i))}{\cosh(\lambda t_f)}\approx\frac{\Delta \mathbf{X}_ie^{\lambda(t-t_i)}}{2}.
\end{equation}
Using the same reasoning for $t>t_i$, each term in the above sum becomes
\begin{eqnarray}
    \Delta \mathbf{X}_i\left(\frac{\sinh(\lambda t)\sinh(\lambda(t_f-t_i))}{\cosh(\lambda t_f)}-\cosh(\lambda(t-t_i))\right)&\approx& \Delta \mathbf{X}_i\left(\frac{e^{\lambda (t-t_i)}}{2}-\cosh(\lambda(t-t_i))\right)\\
    &=&-\frac{\Delta \mathbf{X}_i e^{\lambda (t_i-t)}}{2}.
\end{eqnarray}
Combining these last two equations gives
\begin{equation}
    \mathbf{y}_{dc}(t)\approx\mathbf{X}_{dc}(t)-\sum_i\Delta \mathbf{X}_i \frac{e^{-\lambda |t-t_i|}}{2}\textrm{sgn}(t_i-t). \label{eq:ydcfin}
\end{equation}}

\change{For the continuous part, one has
\begin{equation}
    \mathbf{y}_c(t)=\frac{\mathbf{x}\cosh\left(\lambda(t_f-t)\right)}{\cosh(\lambda t_f)}+\frac{\lambda\sinh(\lambda t)}{\cosh(\lambda t_f)}\int^{t_f}_0d\tau\,\cosh(\lambda(t_f-\tau))\mathbf{X}_c(\tau)-\lambda\int^{t}_0d\tau\,\sinh(\lambda(t-\tau))\mathbf{X}_c(\tau).
\end{equation}
Once again, one can verify that
\begin{equation}
    \cosh(\lambda t_f)\approx \frac{e^{\lambda t_f}}{2},
\end{equation}
as $\lambda$ is large.
Furthermore, one can verify that the first term in the above equation is zero unless $t\approx 0$, in which case
\begin{equation}
    \frac{\cosh(\lambda(t_f-t))}{\cosh(\lambda t_f)}\approx e^{-\lambda t}.
\end{equation}
}
 \change{These approximations lead to}
\change{\begin{equation}
    \mathbf{y}_c(\mathbf{x},t)\sim \mathbf{x}e^{-\lambda t}+2\lambda e^{-\lambda t_f}\sinh\left(\lambda t\right)\int^{t_f}_0d\tau\,\cosh(\lambda\tau)\mathbf{X}_c(t_f-\tau)-\lambda\int^t_0d\tau\,\sinh\left(\lambda\tau\right)\mathbf{X}_c(t-\tau).
\end{equation}}
One can also write
\change{\begin{eqnarray}
     2\lambda e^{-\lambda t_f}\sinh\left(\lambda t\right)\int^{t_f}_0d\tau\,\cosh(\lambda\tau)\mathbf{X}_c(t_f-\tau)&\approx&\frac{\lambda}{2}\int^{t_f}_0d\tau\,\left(e^{\lambda(\tau+t-t_f)}-e^{\lambda(\tau-t-t_f)}+e^{\lambda(-\tau+t-t_f)}\right)\mathbf{X}_c(t_f-\tau)\nonumber\\
     &=&\frac{\lambda}{2}\left(\int^{t}_{-(t_f-t)}d\tau'\,e^{\lambda\tau'}\mathbf{X}_c(t-\tau')-\int^{-t}_{-(t_f+t)}d\tau'\,e^{\lambda\tau'}\mathbf{X}_c(-t-\tau')\right)\nonumber\\
     &&+\frac{\lambda}{2}\int^{t-t_f}_{t-2t_f}d\tau'\,e^{\lambda\tau'}\mathbf{X}_c(2t_f+\tau'-t)\nonumber\\
     &\approx&\frac{\lambda}{2}\int^{t}_{-(t_f-t)}d\tau'\,e^{\lambda\tau'}\mathbf{X}_c(t-\tau')+\frac{\mathbf{X}_c(0)}{2}e^{-\lambda t}+\frac{\mathbf{X}_c(t_f)}{2}e^{-\lambda(t_f-t)},
\end{eqnarray}}
\change{where I used in the first line that the fact that $\exp(-\lambda t_f-\lambda t-\lambda\tau)\approx 0$ for all choices of $t$ and $\tau$. For the third equality I used the fact that the integrandum in the second and third term are always approximately zero appart from a small region around the upper bound of the integrandum. As this region is small, $\mathbf{X}_c$ is approximately constant over this region and the integrandum can be approximated to be an exponential.
Using the same reasoning, one has,}
\begin{eqnarray}
    \lambda\int^t_0d\tau\,\sinh\left(\lambda\tau\right)\mathbf{X}_c(t-\tau)&=&\frac{\lambda}{2}\int^{t}_{0}d\tau\,\left(e^{\lambda\tau}-e^{-\lambda\tau}\right)\mathbf{X}_c(t-\tau)\nonumber\\
    &\approx&\frac{\lambda}{2}\int^{t}_{0}d\tau\,e^{\lambda\tau}\mathbf{X}_c(t-\tau)+\frac{\mathbf{X}_c(t)}{2}-\frac{\mathbf{X}_c(0)e^{-\lambda t}}{2}
\end{eqnarray}
Combining these equations gives
\change{\begin{eqnarray}
     2\lambda e^{-\lambda t_f}\sinh\left(\lambda t\right)\int^{t_f}_0d\tau\,\cosh(\lambda\tau)\mathbf{X}_c(t_f-\tau)- \lambda\int^t_0d\tau\,\sinh\left(\lambda\tau\right)\mathbf{X}_c(t-\tau)&\approx&\frac{\lambda}{2}\int^{0}_{-(t_f-t)}d\tau\,e^{\lambda\tau}\mathbf{X}_c(t-\tau)\nonumber\\&&+\frac{\mathbf{X}_c(t)}{2}-\mathbf{X}(0)e^{-\lambda t}\nonumber\\&&+\frac{\mathbf{X}_c(t_f)}{2}e^{-\lambda(t_f-t)}
\end{eqnarray}}
\change{Furthermore, using the same reasoning as before, one has
\begin{eqnarray}
    \frac{\lambda}{2}\int^{0}_{-(t_f-t)}d\tau e^{\lambda\tau}\mathbf{X}_c(t-\tau)&=&\frac{\lambda}{2}\int^{0}_{-\infty}d\tau e^{\lambda\tau}\mathbf{X}_c(t-\tau)-\frac{\lambda}{2}\int^{-(t_f-t)}_{-\infty}d\tau e^{\lambda\tau}\mathbf{X}_c(t-\tau)\nonumber\\
    &\approx&\frac{\mathbf{X}_c(t)}{2}-\frac{\mathbf{X}_c(t_f)}{2}e^{-\lambda(t_f-t)}.
\end{eqnarray}}
\change{These last two equations lead to
\begin{equation}
  2\lambda e^{-\lambda t_f}\sinh\left(\lambda t\right)\int^{t_f}_0d\tau\,\cosh(\lambda\tau)\mathbf{X}_c(t_f-\tau)- \lambda\int^t_0d\tau\,\sinh\left(\lambda\tau\right)\mathbf{X}_c(t-\tau)\approx\mathbf{X}_c(t)-\mathbf{X}_c(0)e^{-\lambda t}.
\end{equation}}
Plugging this back in into the previous equation, gives
\change{\begin{equation}
    \mathbf{y}_c(x,t)\sim (\mathbf{x}-\mathbf{X}_c(0))e^{-\lambda t}+\mathbf{X}_c(t).~\label{eq:ycfin}
\end{equation}
Combining with Eqs.~\eqref{eq:ydcfin} and \eqref{eq:ycfin} leads to
\begin{equation}
    \mathbf{y}(\mathbf{x},t)\sim\mathbf{X}(t)+(\mathbf{x}-\mathbf{X}_c(0))e^{-\lambda t}+\sum_i\Delta \mathbf{X}_i \frac{e^{-\lambda |t-t_i|}}{2}\textrm{sgn}(t_i-t).
\end{equation}
Plugging this into the expressions of $\epsilon$ and $\Delta_iS$ gives
\begin{eqnarray}
    \epsilon&\sim&\int^{\infty}_{-\infty}d\mathbf{x}\,\int^{t_f}_0dt\,p(x,0)\left(\left(\mathbf{x}-\mathbf{X}(0)\right)e^{-\lambda t}+\sum_i\Delta \mathbf{X}_i\frac{e^{-\lambda\left|t-t_i\right|}}{2}\textrm{sgn}(t_i-t)\right)^2\nonumber\\
    &\sim&\frac{1}{2\lambda}\int^{\infty}_{-\infty}d\mathbf{x}\,p(x,0)\left(\mathbf{x}-\mathbf{X}(0)\right)^2+\frac{\sum_i\Delta\mathbf{X}_i}{\lambda}\nonumber\\
    &=&\frac{\left\langle \left(x-X(0)\right)^2\right\rangle_0+\sum_i\left(\Delta\mathbf{X}_i\right)^2}{2\lambda}\label{eq:eps1},
\end{eqnarray}
and
\begin{eqnarray}
    \Delta_iS&\sim&\frac{k_B}{D}\int^{\infty}_{-\infty}d\mathbf{x}\,\int^{t_f}_0dt\,p(\mathbf{x},0)\left(-\lambda\left(\mathbf{x}-\mathbf{X}(0)\right)e^{-\lambda t}+\frac{\partial}{\partial t}\mathbf{X}_c(t)-\sum_i\frac{\textrm{sgn}(t-t_i)e^{-\lambda\left|t-t_i\right|}}{2}\right)^2\nonumber\\
    &\sim&\frac{k_B\lambda}{2D}\left(\left\langle \left(\mathbf{x}-\mathbf{X}(0)\right)^2\right\rangle_0+\sum_i\left(\Delta\mathbf{X}_i\right)^2\right)\label{eq:dis1},
\end{eqnarray}
which leads to Eq.~\eqref{eq:diseps}}.

\section{Applications\label{app:ec}}
\begin{figure}
    \centering
    \includegraphics[scale=0.7]{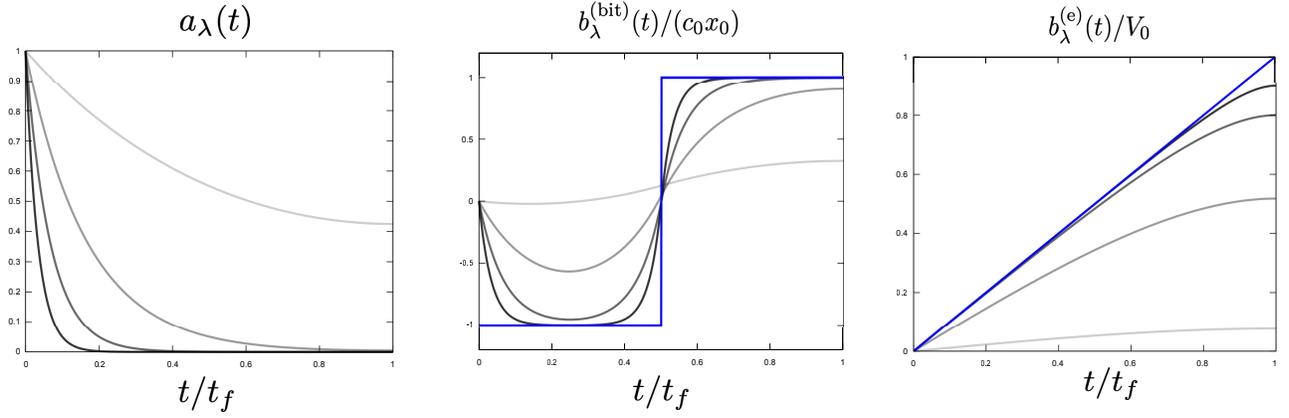}
    \caption{\change{Illustration of $a_{\lambda}(t)$ and $b_{\lambda}(t)$ for bit operations and for the electronic system, cf.~Eqs.~\eqref{eq:bsol1}-\eqref{eq:bsol2}, for $\lambda t_f=0.5,2,5,10$ in increasingly dark shade and $X(t)$ in blue.}}
    \label{fig:aandb}
\end{figure}
In this appendix, I will calculate explicit expressions for the Pareto front under full control, Eqs.~\eqref{eq:sigy}-\eqref{eq:yres0}, \change{for the applications discussed in the main text}. \change{Both bit-erasure and the electronic circuit are one-dimensional systems, with $X(t)$ given by Eqs.~\eqref{eq:Xtinf} and \eqref{eq:Xelec} respectively.}

\change{For the example of bit erasure, one can get an explicit solution for $b_\lambda$, using mathematical software such as Mathematica
\begin{eqnarray}
    b_\lambda(t)&=&c_0x_0\left(\frac{1}{2}+\cosh(\lambda t)-\frac{3}{2}\cosh\left(\lambda\left(t-\frac{ t_f}{2}\right)\right)+\frac{\sinh\left(\lambda t\right)}{\cosh\left(\lambda t_f\right)}\left(2\sinh\left(\frac{\lambda t_f}{2}\right)-\sinh(\lambda t_f)\right)\right)\nonumber\\&&+\left(\cosh\left(\lambda \left(t-\frac{t_f}{2}\right)\right)-1\right)X(t).\label{eq:bsol1}
\end{eqnarray}
One can plug this equation into Eqs.~\eqref{eq:sigy}-\eqref{eq:yres0} and execute the remaining integrals to arrive at a closed parametric equation for the Pareto front under optimal control. These final expressions are very lengthy and will not be shown here.}

\change{One can obtain a similar result for the electronic circuit. Firstly, one can verify that
\begin{equation}
    b_\lambda(t)=V_T(t)-\frac{V_0\sinh(\lambda t)}{\lambda t_f\cosh(\lambda t_f)}.\label{eq:bsol2}
\end{equation}
Once again, one can also explicitly solve the integrals in Eqs.~\eqref{eq:sigy}-\eqref{eq:yres0}. In this case the resulting equations are more straightforward:
\begin{eqnarray}
    \frac{\epsilon}{t_f}&=&\frac{2k_BT(\lambda')^3-2\lambda'CV_0^2+(k_BT(\lambda')^2+CV_0^2)\sinh(2\lambda')}{4C(\lambda')^3\cosh(\lambda')^2},\\
    \frac{t_fT\Delta_iS}{RC}&=&\frac{2\lambda'CV_0^2+(\lambda'CV_0^2-k_BT(\lambda')^3)\textrm{sech}(\lambda')^2+(k_BT\lambda'^2-3CV_0^2)\tanh(\lambda')}{2\lambda'},
\end{eqnarray}
where I introduced $\lambda'=\lambda t_f$ for notational simplicity. One can find the Pareto front by inverting one of these two equations with respect to $\lambda'$ and plugging it in into the other equation. This can generally not be done analytically.
}

\change{One can verify that in both examples for increasing $\lambda$, $b_\lambda(t)$ is increasingly close to the pre-defined trajectory. This is shown in Fig.~4.}

\section{Limited control \label{app:lc}}
The precision associated with the voltage is given by
\begin{eqnarray}
    \epsilon&=&\int^{t_f}_0dt\,\int^{\infty}_{-\infty}dv_c\,\left(v_c-V_T(t)\right)^2p(v_c,t)\nonumber\\
    &=&\int^{t_f}_0dt\,\left(\left(V_C(t)-V_T(t)\right)^2+\frac{k_BT}{C}\right).\label{eq:appeps}
\end{eqnarray}
\change{To calculate the entropy production, one can use Eq.~\eqref{eq:pf} to note that
\begin{equation}
    v(v_c,t)=\frac{V_s(t)-V_C(t)}{RC},
\end{equation}
and
\begin{equation}
    V_S(t)=V_C(t)+RC\frac{\partial}{\partial t}V_C(t),\label{eq:VsVcrel}
\end{equation}
as can be deduced from Eq.~\eqref{eq:vcres}
Therefore,
\begin{eqnarray}
    \Delta_iS&=&-\int^{t_f}_0dt\int^{\infty}_{-\infty}dv_c\,\frac{(V_C(t)-V_s(t))^2}{RT}p(v_c,t)\nonumber\\
    &=&\int^{t_f}_0dt\,\frac{RC^2}{T}\left(\frac{\partial}{\partial t}V_C(t)\right)^2.\label{eq:sapp}
\end{eqnarray}}
Therefore, the Lagrangian is given by
\begin{equation}
    \mathcal{L}=\int^{t_f}_0dt\,\left(\left(V_C(t)-V_T(t)\right)^2+\frac{k_BT}{C}+\frac{1}{\lambda^2}\left(\frac{\partial}{\partial t}V_C(t)\right)^2\right).
\end{equation}
Eq.~\eqref{eq:VsVcrel} shows that minimizing the above Lagrangian with respect to $V_S(t)$ is essentially equivalent to minimizing it with respect to $V_C(t)$. This latter minimization can be done using Lagrangian methods and gives
\change{\begin{equation}
    \frac{\partial^2}{\partial t^2}V_C(t)=\lambda^2\left(V_C(t)-V_T(t)\right),
\end{equation}
or using the fact that $V_T(t)=V_0t/t_f$,
\begin{equation}
    V_C(t)=c_1\sinh\left(\lambda t\right)+c_2\cosh\left(\lambda t\right)+\frac{V_0 t}{t_f},
\end{equation}
where $c_1$ and $c_2$ are integration constants that can be determined through $V_C(0)=0$ (or $c_2=0$), and
\begin{equation}
   \frac{\partial}{\partial c_1}\mathcal{L}=0.
\end{equation}
Solving this last equation gives
\begin{equation}
    c_1=-\frac{V_0}{\lambda t_f\cosh(\lambda t_f)},
\end{equation}
or
\begin{equation}
    V_C(t)=\frac{V_0 t}{t_f}-\frac{V_0\sinh(\lambda t)}{\lambda t_f\cosh(\lambda t_f)},
\end{equation}
}
Filling this back in into Eqs.~\eqref{eq:appeps} and \eqref{eq:sapp} gives a parametric equation for the Pareto front under limited control,
\change{\begin{eqnarray}
    \epsilon&=&\frac{k_BTt_f}{C}+\frac{t_fV_0^2\left(\sinh(2\lambda')-2\lambda'\right)}{4(\lambda')^3\cosh^2(\lambda')},\\
    \Delta_iS&=&\frac{V_0^2\left(2\cosh^2(\lambda')+1\right)-3\sinh(\lambda')\cosh(\lambda')}{2\lambda't_f\cosh^2(\lambda')}.
\end{eqnarray}
}

\end{document}